\documentclass[twocolumn,aps,groupedaddress,nopacs,footinbib,preprintnumbers,floats,floatfix]{revtex4}

\renewcommand{\d}{\mathrm{d}}
\usepackage{color}
\usepackage{amsmath}
\usepackage{amssymb}
\usepackage{latexsym}
\usepackage{graphicx}
\usepackage{subfigure,graphics,epsfig}

\begin{document}

\title{Non-parametric Dark Energy Degeneracies}

\author{Ren\'ee Hlozek$^{1,3}$, Marina Cort\^{e}s$^{1,2}$, Chris Clarkson$^1$ and Bruce Bassett$^{1,3}$\\
\it $^1$Cosmology \& Gravity Group, Department Mathematics
and Applied Mathematics, University of Cape Town, Rondebosch 7701,
South Africa.\\ $^2$Astronomy Centre, University of Sussex, Brighton BN1 9QH, United
Kingdom\\$^3$ South African Astronomical Observatory, Observatory, Cape Town, South Africa}

\begin{abstract}
We study the degeneracies between dark energy dynamics, dark matter and curvature using a non-parametric and non-perturbative approach.
This allows us to examine the knock-on bias induced in the reconstructed dark energy equation of state, $w(z)$, when there is a bias in the cosmic curvature or dark matter content, without relying on any specific parameterisation of $w$. Even assuming perfect Hubble, distance and volume measurements, we show that for $z > 1$, the bias in $w(z)$ is up to two orders of magnitude larger than the corresponding errors in $\Omega_k$ or $\Omega_m$. This highlights the importance of obtaining unbiased estimators of all cosmic parameters in the hunt for dark energy dynamics.
\end{abstract}

\maketitle

\section{\textbf{Introduction}}

Since 1998 \cite{perlm, riess} evidence has been mounting in support of an accelerated expansion of the Universe. Nearly 10 years on, the puzzle of the origin of this acceleration -- dubbed dark energy -- remains one of the most intriguing enigmas in modern day science. Much activity has come from both the theoretical and observational sectors of the physics community in an attempt to pin down its origin.
The current drive in dark energy studies is focused on trying to establish its dynamical behaviour as a function of redshift, $w(z)$. While the simplest explanation remains a $\Lambda$CDM universe with $w=-1$ for all redshift, dynamics in $w(z)$ would provide a window into new physics.
Therefore, uncovering the dynamics of dark energy as described by the ratio of its pressure to density, $w(z) = p_{DE}/\rho_{DE}$, has become the focus of multi-billion dollar proposed experiments using a wide variety of methods, with several planned surveys at redshifts above unity, as high-redshift measurements are useful to constrain dark energy parameters and test for deviation from the concordance $\Lambda$CDM model (see e.g. \cite{DETF}). Unfortunately the search for dynamical behaviour in $w$ is a mani-fold problem. The nature of dark energy is elusive: cosmic observations depend not only on dark energy but also on other cosmic parameters such as the cosmic curvature, $\Omega_k$, and the total matter content, $\Omega_m$, leading to degeneracies between these and $w(z)$ parameters, an issue which has recently been under intense scrutiny by the community \cite{m1, CCB,linder, huang}. Kunz \cite{m1} argues that observations are only sensitive to the full energy-momentum tensor and thus cannot see beyond a combination of the ``dark component" -- dark matter \textit{plus} dark energy. The degeneracy between the geometry of the universe and the equation of state of dark energy has also been discussed in light of the well-known result that a cosmological constant in the presence of spatial curvature can mimic a dynamical dark energy \cite{CCB}.

In this work we review current constraints on cosmic curvature and extend the approach in \cite{CCB} in two ways. First we include the reconstruction of $w(z)$ that would follow from measurements of the rate of change of cosmic volume with redshift, $dV/dz$. Secondly we discuss the $w(z)$ that would be incorrectly reconstructed from perfect Hubble, distance and volume data if the wrong value of $\Omega_m$ were used. We assume perfect data for all observations, which allows us to probe fundamental, ``in-principle" degeneracies that are not due to finite errors and incomplete redshift-coverage. This implies that given a specific bias in a cosmological parameter, the degeneracies will be true no matter what progress is made in improving future cosmic surveys. Furthermore, the key point in our reconstruction of $w(z)$ is that it is performed in a fully non-parametric manner, and so does not rely on the validity of any particular parameterisation of $w(z)$. To illustrate the power of this  non-parametric approach, we compare our method with a standard equation of state parameterisation \cite{cp,lin}, which cannot fully resolve the above degeneracies.

\subsection{Degeneracies in Dark Energy Studies}

The success of the inflationary scenario for the early Universe and its standard prediction of flatness to high precision ($\Omega_k < 10^{-10}$) is perhaps the main reason why curvature has traditionally been left out in analyses of dark energy. However, possible scenarios in which inflation is consistent with non-zero spatial curvature have recently been investigated \cite{freivogel}. It is also interesting to note that the backreaction of cosmological fluctuations may cause effective non-zero curvature that may yield practical limits on our ability to measure $w(z)$ accurately at $z > 1$ (see e.g. \cite{backreact}).
 Since measurements of the Cosmic Microwave Background (CMB) have so far proved consistent with flatness (e.g. \cite{wmap3}) statistical quantities that measure the necessity of introducing extra parameters (such as Bayesian Evidence or information criteria \cite{BCK, m2, trotta}) do not favour the inclusion of curvature as a parameter in current analyses \cite{liddle2}. However this is more a symptom of the inability of current data to constrain an extra parameter than a conclusive case for a flat universe. Further, Bayesian evidence or information criteria do not take into account the power of the biases that may be introduced by falsely neglecting a parameter. We will show below that the biases introduced in neglecting curvature are very significant at $z > 1$.

In general constraints on curvature are very fragile to assumptions about the dark energy since they are primarily derived from distance measurements ($d_L$ or $d_A$) which are completely degenerate with curvature \cite{weinberg70}. One way to illustrate the degeneracy between curvature and dynamics is as follows. Let us assume that we know all cosmic parameters perfectly other than the curvature $\Omega_k$ and the dark energy equation of state, $w(z)$. At any redshift, $z_*$, a perfect measurement of $d_L(z_*)$ (or $d_A(z_*)$) allows us to measure a single quantity. If we know $w(z_*)$ then that quantity can be $\Omega_k$. However, if $w(z)$ is truly a free function, then its value at $z_*$ is completely free and we are left trying to find two numbers from a single observation, which is impossible.

Only when we start to correlate the values of $w(z)$ at different redshifts can we begin to use distance measurements alone to constrain the curvature. The standard way to do this is to assume that $w(z)$ can be compressed onto a finite-dimensional subspace described by $n$
parameters, e.g.
\begin{equation}
w(z) =  \Sigma_j^n w_j z^j
\end{equation}
In this case perfect distance measurements at $n+1$ different redshifts will allow a complete solution of the problem and will yield the $w_j$ and $\Omega_k$. The most extreme version of this is to assume $\Lambda$CDM, $w(z)=-1$. Within this context it is of course possible
to derive very stringent constraints on the curvature. For example, combining the WMAP 3 year data and the SDSS DR5 Luminous Red Galaxy (LRG)  sample leads to $\Omega_k = -0.003\pm 0.010$ assuming $w = -1$
\cite{sdsslrg}. The addition of extra data is crucial since the WMAP data alone provides
only the constraint $\Omega_k = -0.3040 + 0.4067 \Omega_{\Lambda}$
\cite{wmap3}.

\begin{figure}
\begin{center}
\includegraphics{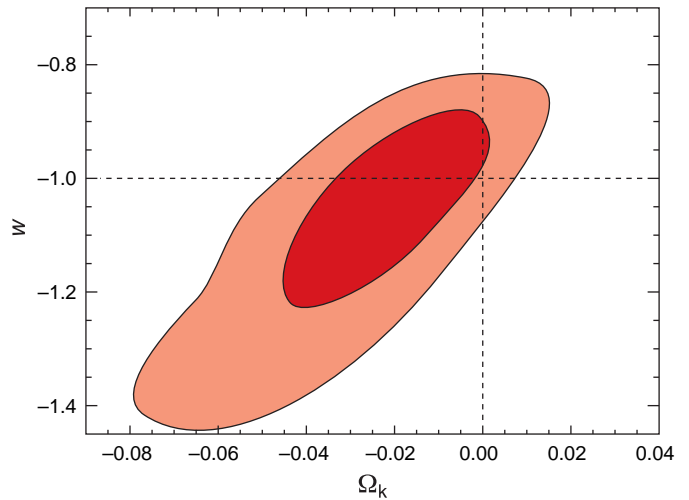}
\caption{{\bf The curvature-dark energy degeneracy} Contours showing the 2D marginalized contours for $w$ and $\Omega_k$ based on combined data from WMAP3, 2dFGRS, SDSS and supernova surveys. While the slope of the degeneracy differs for this combination of data, the sign of the degeneracy is consistent with the $w_0$ term in Eqs. (\ref{degenhub}),(\ref{degend}). Taken from \cite{wmap3}.\label{wmap17}}
\end{center}
\end{figure}

It is a highly non-trivial statement that flat $\Lambda$CDM models provide such a good fit to all the data, but we must be aware that such constraints on the curvature are artificially strong in the sense that adding more dark energy parameters will lead to an almost perfect
degeneracy with the curvature. This is visible in Fig. 17 of the WMAP3 \cite{wmap3} (here Fig.~\ref{wmap17}) which shows the correlation between a constant $w$ and $\Omega_k$.

Hence we can currently say very little about the true value of the
$\Omega_k$ and the belief that the spatial curvature is small
is essentially based on Occam's Razor. Although one could fit distance
measurements with any value of $\Omega_k$, the required $w(z)$
functions would be disfavoured by Bayesian model selection which
penalize models with extra parameters that do not significantly
improve the fit. We show in detail later the required $w(z)$ functions
to do precisely this.

At present a well-defined program for measuring the spatial curvature
of the cosmos does not exist. To illustrate this, consider fixing the
dark energy to be described by only $n$
parameters. One would hope that given this restriction the resulting
constraints on $\Omega_k$ would be independent of the precise choice
of the $n$ parameters, i.e. independent of the parameterisation.
Unfortunately a little thought makes it clear that this cannot be
true. A parameterisation of $w(z)$ which does not allow mimicry of
curvature will provide good, decorrelated constraints on the curvature
(which does not mean
the corresponding best-fit will be a good fit to the data) while a
model which allows perfect mimicry of the dynamics of the curvature
(i.e. $(1+z)^2$) will show highly correlated constraints (although for
negative $\Omega_k$ mimicry of distance data is only possible up to a
critical redshift as we show later).

\begin{figure}[htbp!]
\begin{center}
\includegraphics[width=0.5\textwidth]{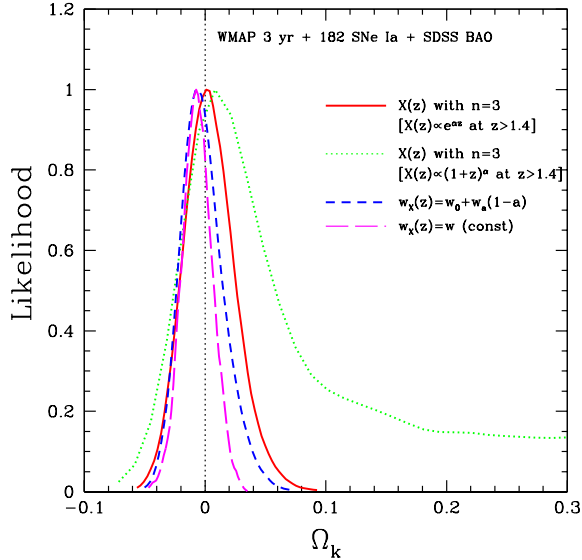}
\caption{{\bf The curvature-dark energy degeneracy} likelihood for $\Omega_k$ for different parameterisations of dark energy. Assuming a constant model for $w$, allows $\Omega_k$ to be tightly constrained at $2\sigma$ to be near 0. However introducing dynamics reduces these constraints significantly. Here $X(z) = \rho_X(z)/\rho_X(0)$ is the dark energy density, which \cite{wang} assume is a free function below some cut-off redshift $z_{cut}$. The value of $X$ at redshifts $z_i = z_{cut}(i/n), i = 1,2..,n$ are treated as $n$ independent model parameters that are estimated from the data. A specific functional form for $X$ is assumed above the cut-off redshift. The likelihoods are given for two such forms of $X(z)$;  namely a power law, $X \propto (1+z)^{\alpha}$ for $ z > z_{cut} $, and an exponential function $X \propto e^{\alpha z}$. In these figures there are $n = 3$ independent redshifts below a cut-off redshift of $z_{cut} = 1.4$. Taken from \cite{wang}\label{wangfig}.}
\end{center}
\end{figure}

\begin{figure*}[htbp!]
\centering
\mbox{
     \subfigure{\scalebox{0.8}{\includegraphics{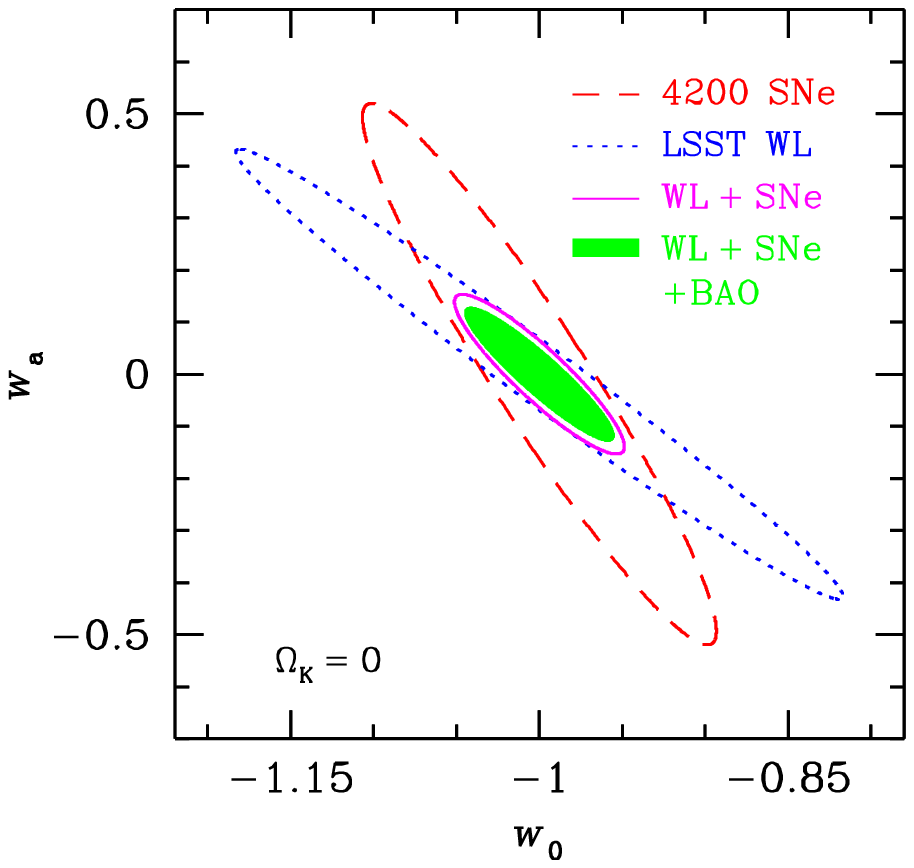}}} \quad
     \subfigure{\scalebox{0.8}{\includegraphics{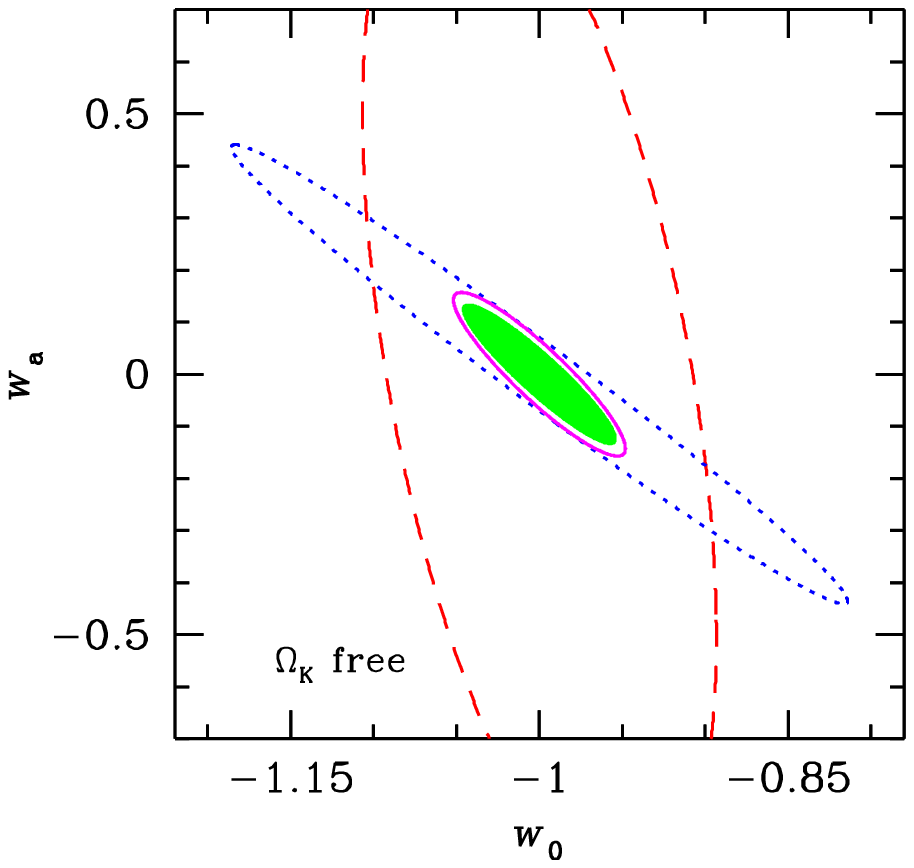}}}}
\caption{{\bf Left} - $1\sigma$ error contours assuming flatness for the dark energy parameters $w_0$ and $w_a$ for the CPL parameterisations. {\bf Right} - as on the left but with curvature left free and marginalised over. Note how
pure distance measurements suffer strongly even with the very limited $w(z)$
parameterisation but that when all the surveys are combined the final error
ellipse is essentially unaffected. This is to be expected from Equation (\ref{curv})
which shows how $\Omega_k$ can be determined from simultaneous Hubble rate
and distance measurements. Figure from Knox {\em et al.} \cite{knox}.\label{knoxfig}}
\end{figure*}

This dilemma is visible in various recent studies attempting to
constrain cosmic curvature in the presence of multiple dark energy
$w(z)$ parameters
\cite{groupbb, wang, huang}. For some popular parameterisation constraints on $\Omega_k$ are of
order $|\Omega_k| < 0.05$ at $2\sigma$. For other parameterisation the
constraints evaporate and
even $\Omega_k \sim 0.2$ cannot be ruled out (see Fig.~\ref{wangfig}, which is taken from \cite{wang}).

Alternative cosmic measurements sensitive to curvature include the Integrated Sachs Wolfe (ISW) effect, which is sensitive to the growth of the metric fluctuations $\Phi$, which is in turn sensitive to both dark energy and curvature. Recent work to investigate the ISW effect as a function of redshift uses the combination of CMB data with information on large scale structure \cite{isw}. Combining WMAP with such suitable tracers of large scale structure shows that $\Phi$ has been decreasing with cosmic time \cite{nolte}, which rules out a large positive curvature which would have predicted the opposite trend.

Another measurement sensitive to the growth function is differential number counts $dN/dz$, e.g. of clusters. This is a potentially sensitive test which, given a constant comoving number of objects, reduces to a test of the rate of change of cosmic volume with redshift, $dV/dz$. We discuss in detail below how perfect measurements of $dV/dz$ allow reconstruction of $w(z)$, and we discuss the resulting errors on dark energy when systematic biases in cosmic parameters are present.

Measurements of the power spectrum from CMB data and from measurement of Baryon Acoustic Oscillations (BAO) provide estimates of the matter content of the universe. While constraints on $\Omega_m$ are sharpened by combining data from many observations, the best-fit value is often derived on the assumption of flatness \cite{sdsslrg, perciv}. Unlike the case for cosmic curvature the degeneracy between observables and the matter content is perfect and we show that incorrectly assuming a particular value for $\Omega_m$ can also mimic deviations from $\Lambda$CDM.

\subsection{Future surveys}

We will show in equation (\ref{curv}) that simultaneous measurements of the Hubble
rate $H(z)$, distance $D \propto d_A,d_L$ and $D'(z)$ allow for a perfect
measurement of $\Omega_k$. BAO allow for the
simultaneous measurement of both distance and Hubble rate at the central
redshift \cite{bao}. For a flat universe $D'(z) \propto 1/H(z)$, but in a
curved universe this is not true: the curved geodesics mean that $D'(z)$
contains extra information encoded in $\Omega_k$.

Measuring $D'(z)$ is in principle possible with future BAO, weak lensing and
supernova surveys. In particular, cross-correlation tomography of deep
lensing surveys appears to be a very powerful probe of curvature when
combined with BAO surveys \cite{bernstein}, assuming that self-calibration is
possible. In principle it should be possible to measure the cosmic curvature
to an accuracy of about $\sigma(\Omega_k) \simeq 0.01$ for an all-sky weak
lensing and BAO survey out to $z=10$. In principle such a survey would be
able to measure distances to about $10^{-4} f_{sky}^{-1/2}$ in redshift bins
of width $\Delta z = 0.1$ out to $z=2.5$ \cite{bernstein}. This relies
critically on the combination of weak lensing and BAO data since constraints from either
observations alone are significantly degraded. This can also be seen in Fig. (\ref{knoxfig}) which
shows the error ellipses for the parameters in the CPL parameterisation $w(z)
= w_0 + w_a \left(\frac{z}{1+z}\right)$ assuming flatness (left) and leaving $\Omega_k$ free
(right). Note that although individual error ellipses are significantly degraded, the
combined data sets have an almost unchanged error ellipse.
\newline

Our work is organized as follows: we illustrate the dependence of the background observables on the cosmological parameters $\Omega_m, \Omega_k$ in section \ref{bg_expand} and discuss obtaining the dark energy equation of state $w$ from observables in section \ref{obs_eos}. The process of reconstructing $w$ via a non-parametric approach is described in section \ref{recon_w}. Finally we link this non-parametric approach to other standard approaches to dark energy degeneracies in section \ref{param} and conclude in section \ref{conc}.

\section{Dark Energy from observations \label{de_obs}}

There are three key observables of the background geometry which play a pivotal role in determining $w(z)$, namely measurements of distances, of the expansion history (i.e. the Hubble parameter) and of the change in the fractional volume of the Universe (e.g. from number-counts).

The principle method to date is to relate measurements of the distances of objects to the cosmology of the Universe. This is done via either standard `rulers' of known length - giving the angular diameter distance $d_A(z) $ - or via standard `candles' of known brightness which results in the luminosity distance $d_L(z)$. These are related via the reciprocity relation $d_L(z)=(1+z)^2d_A$. In an FLRW model, these are given by $d_L(z)=c(1+z)D(z)/H_0$, where we define
\begin{equation}
\label{d_L}
D(z)=\frac{1}{\sqrt{-\Omega_k}}\sin{\left(
\sqrt{-\Omega_k}\int_0^z{\mathrm{d}z'\frac{H_0}{H(z')}}\right)}.
\end{equation}
Here, $\Omega_k$ is the usual curvature parameter, and $H(z)$ is given by the Friedmann equation,
\begin{equation}
H(z)^2= H_0^2\left[\Omega_{m} (1+z)^3+\Omega_{k} (1+z)^2+\Omega_{\mathrm DE} f(z)\right]
\label{hubble}
\end{equation}
where
\begin{equation}
\label{f}
f(z)=\exp{\left[3\int_0^z\frac{1+w(z')}{1+z'}\mathrm{d}z'\right]}
\end{equation}
and $\Omega_{DE}=1-\Omega_m-\Omega_k$. Thus, given a cosmological model, we may calculate any distance measure we choose.

The Hubble parameter is in itself an observable which will play an important role in future dark energy experiments. Knowledge of $H(z)$ allows us to directly probe the dynamical behavior of the universe, and it will be directly determined from BAO surveys which simultaneously provide the angular diameter distance, $d_A$ at the same redshift by exploiting
the radial and angular views of the acoustic oscillation scale \cite{bao}, a fact that
will provide key new data in coming years \cite{sdsslrg, wfmos, wigglez}.

The third key background test we will discuss here is the observation of fractional volume change as a function of redshift,
\begin{equation}
\label{wvol}
{V}'(z)\equiv\frac{\d^2 V}{\d z\d\Omega}= \frac{c^3D(z)^2}{H_0^2 H(z)},
\end{equation}
which can in principle be determined via number-counts or the BAO.

Given any two of the above observables we may deduce the third. Perfect observations of these observables should allow us, in principle, to be able to reconstruct two free \emph{functions} when in fact we only need to reconstruct one, namely $w(z)$, as well as two cosmological parameters, $\Omega_m$ and $\Omega_k$.  (Note that if we know $H(z)$ perfectly, we know $H_0=H(0)$, and so this is no longer a free parameter in the same sense.) How do we find these?

We may determine the curvature directly, and independently of the other parameters or dark energy model via the relation~\cite{CCB}
\begin{equation}\label{curv}
\Omega_k=\frac{\left[H(z)D'(z)\right]^2-H_0^2}{[H_0D(z)]^2},
\end{equation}
which may be derived directly from Eq.~(\ref{d_L}). Such independent measurements of the curvature of the universe can in turn be used to test the Copernican Principle in a model-independent way. \cite{copernican}

\subsection{Expansions of the background observables \label{bg_expand}}

To illustrate the dependency of the background observables we consider here we expand them in terms of the cosmological parameters $\epsilon_m$, $\Omega_k$ and the parameter $x=z/(1+z)$. Here $\epsilon_m := \Omega_{m*} - \Omega_m$, where $\Omega_{m*}$ is the true value of the matter energy density and $\Omega_m$ is the assumed value, as seen in Eq. (\ref{w0expansion}).

The expansions for $H(x), d_L(x), V'(x)$ yield

\begin{widetext}
\begin{equation*}
x = \frac{z}{1+z}
\end{equation*}

\begin{equation}
\label{hx}
H(x)= H_0\bigg[1+\frac{1}{2}\big\{3(1+w_0(1-\Omega_{m*}))x-(1+3w_0)\Omega_k x-3w_0\epsilon_m\big\}\bigg]
\end{equation}

\begin{equation}
\label{dlx}
d_L(x) = \frac{cx}{H_0}\bigg[ 1+ \big\{( 5 + 3w_0(\Omega_{m*}-1))  + (1+3w_0)\Omega_k  + 3w_0 \epsilon_m   \big\}x\bigg]
\end{equation}

\begin{equation}
\label{vx}
V'(x) = \frac{c^3x^2}{H_0^3}\bigg[1 + \big\{(-1+3w_0(\Omega_{m*} -1)) + (1+3w_0 )\Omega_k + 3w_0 \epsilon_m \big\}x\bigg]
\end{equation}

\end{widetext}

It can be seen from equations (\ref{hx}, \ref{dlx}, \ref{vx}) that the leading term corresponds to that of the standard flat $\Lambda$CDM model. From these equations we can directly compute the error on the particular observable as a function of redshift based on the difference between the `true' cosmology and the `assumed' cosmological model.

\subsection{Obtaining the Dark Energy equation of state from Observations \label{obs_eos}}

Assuming we have `perfect' and uncorrelated data from observations we would like to reconstruct $w(z)$ without assuming a specific parameterisation. Depending on the particular observable of interest, there are different ways to reconstruct $w$.
\newline

{\textbf{\emph{ Dark energy from Hubble}}}

It is straightforward to find $w(z)$ from the Hubble rate~\cite{huterer_hzw, linder}, from Eq.~(\ref{hubble}), and is given by:
\begin{equation}\label{wH}
w(z)=-\frac{1}{3}\frac{\Omega_k H_0^2(1+z)^2+2(1+z)HH'-3H^2}{
H_0^2(1+z)^2[\Omega_m(1+z)+\Omega_k]-H^2}.
\end{equation}
\newline
This tells us $w(z)$ \emph{provided} we already know $\Omega_m$ and $\Omega_k$. However, this reveals a degeneracy between $\Omega_m$ and $w(z)$ which cannot be overcome by background tests alone~\cite{m1}. In essence, geometric background tests can measure the combination $\Omega_{m}+\Omega_{DE}f(z)/ (1+z)^3$, but not the two separately. Another way to view this is by differentiating Eq.~(\ref{wH}), and eliminating $\Omega_m$ to give a differential equation for $w(z)$ in terms of $H, H'$ and $H''$; the constant arising in the general solution to this differential equation is $\Omega_m$.

Similarly, we can reconstruct $w(z)$ from the other two tests on their own.

\begin{widetext}

{\textbf{\emph{ Dark energy from distance measurements}}}
\newline

From distance measurements, we may invert Eq.~(\ref{d_L}) to find
\begin{equation}\label{wD}
w \left( z \right) ={\frac {2\, \left( 1+z \right)  \left(
 D ^{2}\Omega_{{k}}+1  \right)  {D}''  - {D}'  \left[ \Omega_{{k} } \left( 1+z \right) ^{2}  {D}'^{2}+2\,\Omega_{{k}} {D}   \left( 1+z \right) {D}' -3-3\,
  {D}^{2}\Omega_{{k}} \right] }
{3 \left\{  \left[ \Omega_{{k}}+\Omega_{{m}} \left( 1+z \right)
 \right]  \left( 1+z \right) ^{2}D'^{2}- D^{2}\Omega_{{k}}-1 \right\}{D}'}}.
\end{equation}

Reconstructing $w(z)$ from volume measurements as an analytical formula is rather tricky (as it involves the root of a quartic power). It is simpler instead to reconstruct $w(z)$ by solving the differential equation for $f(z)$ and then differentiating to get $w(z)$.\\

{\textbf{\emph{ Dark energy from volume measurements}}} \newline

Starting with Eq. (\ref{d_L}), we solve for the derivative of the Hubble parameter and equate this with the expression for $H'$ in terms of $w(z)$ from Eq. (\ref{wH}) and use
\begin{equation}
\label{wfp}
w(z) = \frac{(1+z)f'}{3f}-1
\end{equation}
to yield a first order differential equation for $f$, namely
\begin{equation}
f'(z)= \frac{A(z) + B(z) + C(z)}{-H_0^2 V'\Omega_{\mathrm{ DE}}},
\end{equation}
where

\begin{equation*}
A(z)=-4{\left(V'H_0 \left(c^3 \sqrt{f(z)\Omega_{\mathrm{DE}} + X_{11}} + V'H_0^3 \Omega_k \left( f(z)\Omega_{\mathrm{ DE}}+ X_{11}\right)\right) \right)}^{1/2},
\end{equation*}
with
\begin{equation*}
X_{ a b}= (1+z)^2(a \Omega_k +  b \Omega_m (1+z)),
\end{equation*}
\begin{equation*}
B(z) = 2H_0^2 V''\left(  f(z)\Omega_{\mathrm{DE}}  + X_{11} \right)
\end{equation*}
and
\begin{equation*}
C(z) = H_0^2 V''\frac{X_{32}}{1+z}.
\end{equation*}
We solve this for $f(z)$ and then use (\ref{wfp}) again to yield $w(z)$. The solution for $f(z)$ is unique since we demand $f(0)=1$.\\

\end{widetext}

\section{Reconstructing $w(z)$ \label{recon_w}}

If we knew $\Omega_m$ and $\Omega_k$ perfectly then our three expressions for $w(z)$ would yield the same function $w(z)$, assuming we lived in an exact FLRW universe. But what if -- as is commonly assumed -- we impose $\Omega_k = 0$ when {\em in fact} the true curvature is actually non-zero? It is usually implicitly assumed that
the error on $w(z)$ will be of order $\Omega_k$, but, as was shown in~\cite{CCB} this is not the case. Will measuring $V'(z)$ possibly circumvent this? And furthermore, are there similar issues from an imperfect knowledge of $\Omega_m$?
\begin{figure*}[p!]
\includegraphics[width=0.95\textwidth]{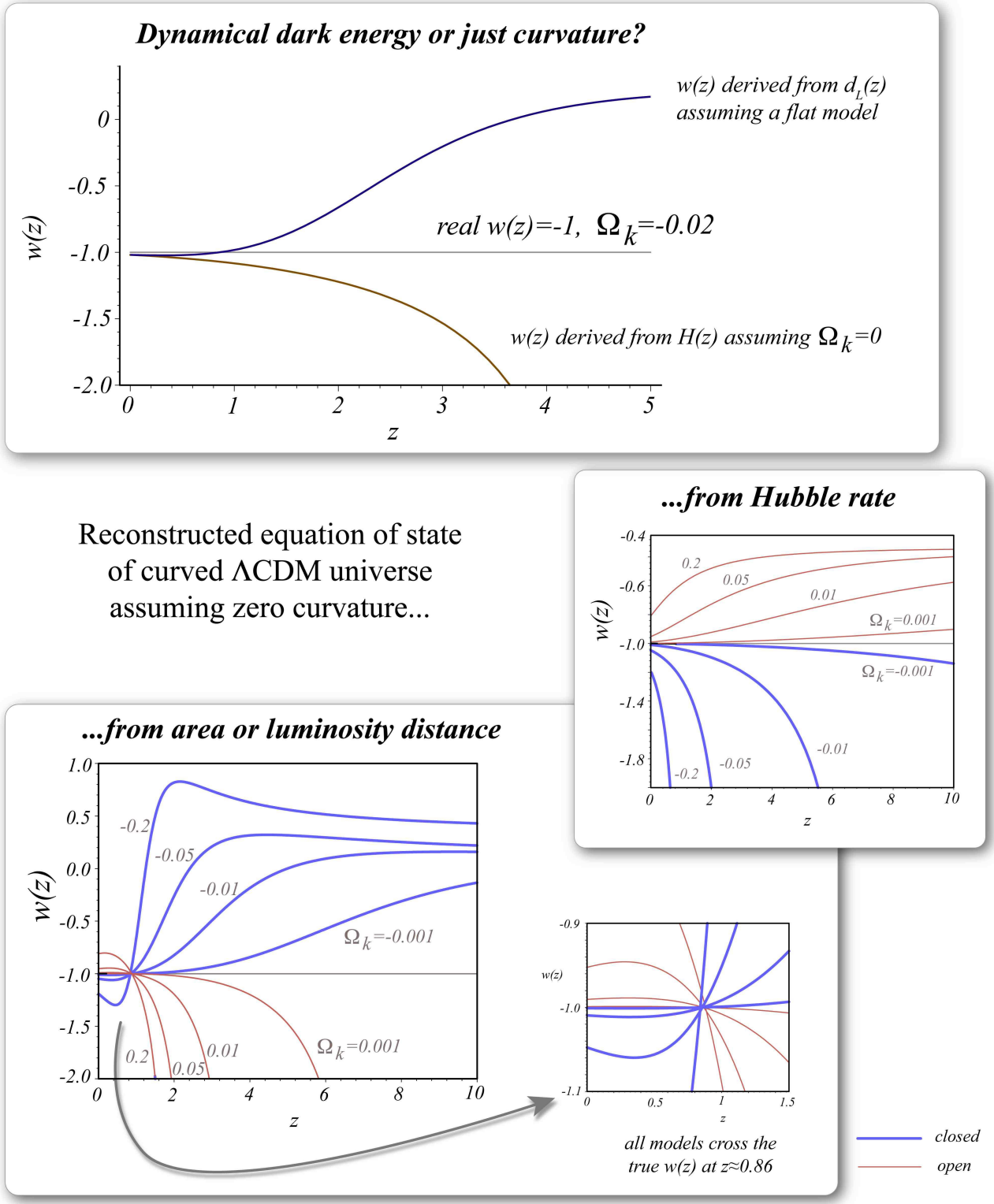}
\caption{\label{ok_dlh}{\bf Reconstructing the dark energy equation of state assuming zero curvature} when the true curvature is 2\% in a closed $\Lambda$CDM universe. The $w(z)$ reconstructed from $H(z)$ is phantom ($w < -1$) and rapidly acquires an error of order 50\% and more at redshift $z\gtrsim2$, and diverges at finite redshift.  The reconstructed $w(z)$ from $d_L(z)$ for $\Omega_k<0$ is phantom until $z\approx0.86$, where it crosses the true value of $-1$ and then crosses 0 at high redshift, where the bending of geodesics takes over from dynamical behavior, producing errors in opposite direction to the DE reconstructed from $H(z)$.
In order to make up for the missing curvature, the reconstructed dark energy is
behaving like a scalar field with a tracking behavior. These effects arise even if the curvature is extremely small $(<0.1\%)$. Reprinted from~\cite{CCB}.\label{ccbfig}}
\end{figure*}
\subsection{{Zero curvature assumption}}
We can easily see the implications of incorrectly assuming flatness by constructing the functions $d_L(z)$ and $H(z)$ under the assumption of the $\Lambda$CDM in a curved Universe (i.e. assuming $w=-1, \Omega_k\neq 0$) and inserting the results into Eqs (\ref{wH}) and (\ref{wD}).

If we then set $\Omega_k=0$ in Eqs. (\ref{wH}) and (\ref{wD}) we arrive at the two corresponding $w(z)$ functions (if they exist) required to reproduce the curved forms for $H(z)$ and $d_L(z)$ in a {\em flat} Universe with dynamic dark energy. This would apply equally to $d_A(z)$ for that matter - the results are exactly the same for any distance indicator. Figure~\ref{ccbfig} presents this method using for simplicity the concordance value of $w=-1$ but we have checked that the qualitative results do not depend on the `true' underlying dark energy model \footnote{In fact, the results presented here are qualitatively the same for any assumed $\Omega_k$ which is different from the true value.}. We assume $\Omega_m=0.3$ in all expressions; numbers quoted are weakly dependent on this. The resulting (spurious) $w(z)$ can then be thought of as the function required to yield the same $H(z)$ or $d_L(z)$ as in the actual curved $\Lambda$CDM model: e.g.,
\begin{equation}
d_L[\text{flat},w(z)]=d_L[\text{curved},w(z)=-1].
\end{equation}

For example for the Hubble rate the reconstructed $w(z)$ can be found analytically to be
\begin{equation}
w(z) = -\frac{1}{3}\frac{\Omega_k (1+z)^2 + 3\Omega_{\mathrm{DE}}} {\Omega_k (1+z)^2 + \Omega_{\mathrm{DE}}},
\end{equation}
without any dependence on a specific parameterisation.

In the figure we show what happens for $\Lambda$CDM: curvature manifests itself as evolving dark energy. In the case of the Hubble rate measurements this is fairly obvious - we are essentially solving the equation $\Omega_{DE} f(z) = \Omega_{\Lambda} + \Omega_k (1+z)^2$ where $f(z)$ is given by Eq.~(\ref{f}). For $\Omega_k > 0$, $w(z)$ must converge to $-1/3$ to compensate for the curvature. For $\Omega_k < 0$, the opposite occurs and a redshift is reached when $w \rightarrow -\infty$ in an attempt to compensate albeit unsuccessfully for the positive curvature. Already we can see why the assumption that the error in $w$ is of order the error in $\Omega_k$ breaks down so drastically.

Interestingly, the curved geodesics imply that the error in $w$ reconstructed from $d_L(z)$ and $H(z)$ have opposing signs at $z \gtrsim 0.9$, as can be seen by comparing the panels for the Hubble rate and the distance indicator in Fig.~\ref{ccbfig}. Above the critical redshift the effect of curvature on the geodesics becomes more important than the pure dynamics, and the luminosity distance flips $w(z)$ in the opposite direction to that reconstructed from $H(z)$.

In the case of volume measurements the reconstructed $w(z)$ has a similar form to the $w$ we obtained from the distance measurements $D(z)$. This can be seen from Eq.~(\ref{wvol}), where the distance information enters the equation as a square power. For example in the closed Universe case the reconstructed $w(z)$ drops to more phantom values ($-2.5$ compared to $-1.3$ for the distance measurements) in order to make up for the missing curvature.

Again the effect of curvature on the geodesics dominates the effect of dynamics for large $z$, and the distance contribution in the volume measurements flips the reconstructed $w(z)$ at $ z = 1.6$. The critical redshift of this flip is determined by the redshift at which the curvature of the geodesics affecting distance measurements becomes more important than the expansion rate. This playoff becomes more finely balanced for volume measurements due to the fact that $H(z)$
appears both in $D(z)$ (as a square power) and on its own. Hence $w(z)$ has to work harder in reproducing curvature to counterbalance the opposing trends of expansion history and
geometry, and so the balance is achieved at higher redshift. The specific redshift at which this happens is dependent on $\Omega_m$ in that lower values imply higher value of the critical redshift.
\begin{figure}[htbp!]
\begin{center}
\includegraphics[width=0.45\textwidth]{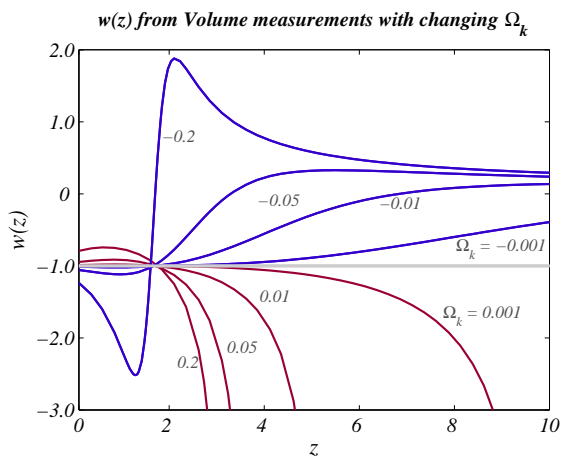}
\caption{\label{okvol} {\bf Reconstructed dark energy from  volume measurements while incorrectly assuming flatness -} Similar to the case for distance measurements in a closed Universe, the reconstructed $w(z)$ must initially be phantom in order to compensate for curvature, and crosses the true value of $w = -1$ at a redshift of $z~1.6$, which is greater than the redshift of 0.86 for the distance measurements alone \cite{CCB}. After this point, the $w(z)$ increases to overcome the curvature of the geodesics.}
\end{center}
\end{figure}

We have shown that incorrectly assuming flatness can result in a reconstructed $w(z)$ that mimics dynamics, yielding errors on $w$ that are much larger than the order of errors on $\Omega_k$. One might then ask if similar errors will result when incorrectly assuming a particular value for the matter density in the Universe, $\Omega_m$.
\subsection{Uncertainties in the Matter content $\Omega_m$}
We consider the similar case of reconstructing $w(z)$ in a flat Universe but here the errors occur when assuming the concordance value of $\Omega_m = 0.3$ incorrectly. For example in this case the $w(z)$ reconstructed from Hubble measurements Eq. (\ref{wH}) reduces to

\begin{equation}\label{wHom}
w(z)=-\frac13\frac{2(1+z)HH'-3H^2}{H_0^2(1+z)^2[\Omega_m(1+z)]-H^2}.
\end{equation}

\begin{figure}[htbp!]
\begin{center}
\includegraphics[width=0.45\textwidth]{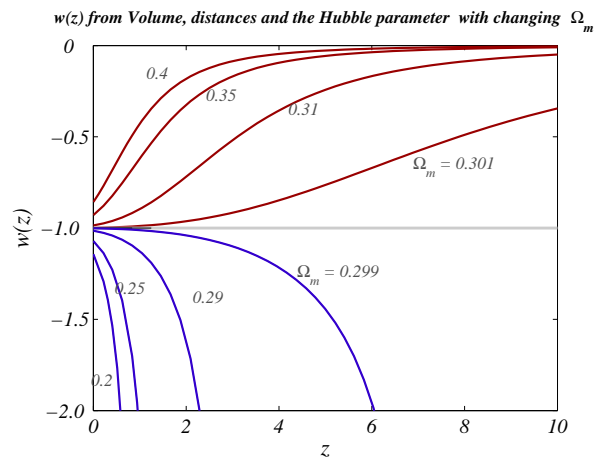}
\caption{ {\bf Reconstructed dark energy from an incorrectly estimated matter density -} The reconstructed $w(z)$ for changing $\Omega_m$ from all three measurements ($H, D, dV/dz$). Since we assume flatness while changing $\Omega_m$, all three observables yield the same reconstructed $w(z)$, since $\Omega_m$ only enters the functions through $H(z)$ or integrals of $1/H$. For $\Omega_m > 0.3$ the dark energy tries to compensate for the extra matter contribution and so asymptotes to $w = 0$ as $ z\rightarrow \infty$. For $\Omega_m < 0.3$ the $w(z)$ is of the same form to what is reconstructed from neglecting curvature in a closed Universe (see Fig.~\ref{ok_dlh}), and the phantom $w$ tends to $-\infty$ as it attempts to compensate for the `missing' matter density.\label{omplot} }
\end{center}
\end{figure}

\begin{figure}[htbp!]
\begin{center}
\includegraphics[width=0.45\textwidth]{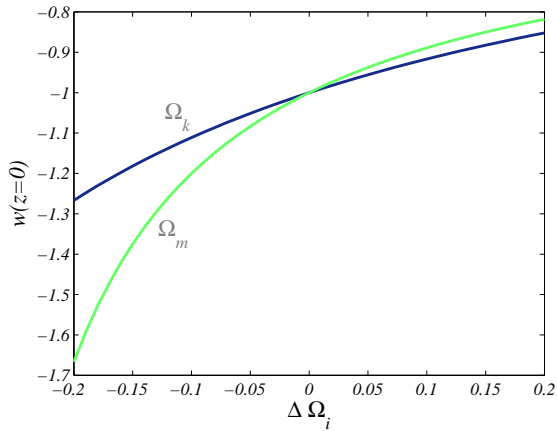}
\caption{ {\bf Low redshift variation in $w(z)$ from $H(z)$ and $D(z)$-}  incorrectly assuming concordance values of $\Omega_m = 0.3$  and $\Omega_k = 0$ results in a variation in the low-redshift value of $w(z)$ reconstructed from observables. The relationship between the error in the cosmological parameter and the reconstructed value for $w$ (while keeping the other cosmological parameter fixed at the prior value) is shown for both $\Omega_m$ (the {\bf green} curve) and $\Omega_k$ (the {\bf blue} curve). \label{diffw} }
\end{center}
\end{figure}

Similar expressions are found for both the distance and volume measurements. The $w(z)$ curves obtained from incorrectly assuming $\Omega_m = 0.3$ are shown in Fig. \ref{omplot}. If we assume flatness for this example we find that changing the value of $\Omega_m$ can only affect the dark energy density, and thus change the value of $H(z)$. As $\Omega_m$ is only present in all three observables through $H(z)$ or integrals of $1/H(z)$, the reconstructed $w(z)$ is the same for all three measurements. Interestingly, the reconstructed $w(z)$ curves do not go through $w = -1$ at $z = 0$, but are spread between -0.85 and -1.15 for $0.2< \Omega_m < 0.4$. This is also shown in Fig.~\ref{diffw}.

Given any scenario of an assumed cosmology that differs from the `true' Universe, we can derive the value of today, $w(z=0)$ from both the Hubble and distance measurements as

\begin{eqnarray}
\label{w0expansion}
w(0) &=&\frac{3-4\Omega_{k*}-3\Omega_m +\Omega_k}{6\Omega_m+6\Omega_{k*} -3\Omega_{m*}-3-3\Omega_{k}} \\
&\sim& \frac{\epsilon_m}{(-1+\Omega_{m*})}-\frac{2\Omega_k}{3(-1+\Omega_{m*})}-1 \nonumber,
\end{eqnarray}

where $\epsilon_m = \Omega_{m*} - \Omega_m$ as defined above where the asterisk indicates assumed but incorrect values of the corresponding quantities. We vary this equation in one `true' density ($\Omega_m $ or $\Omega_k$) at a time, while keeping the other constant at the assumed value of either $\Omega_k = \Omega_{k*}$ or $\Omega_m = \Omega_{m*}$ to produce the curves in Fig.~\ref{diffw}. This parameter $w(z=0)$ allows us to easily quantify the affect of assuming an incorrect cosmological model on the inferred low-redshift value of $w$.

\section{Parametric Degeneracies \label{param}}

We now want to connect the non-parametric approach we have followed above with standard approaches to degeneracies and so we expand Eqs. (\ref{wH}) and (\ref{wD}) for the Hubble rate and distance measurements to first order in $ x = z/(1+z)$. This allows us to link to the parameters $w_0, w_a$ used in the Chevallier-Polarski-Linder (CPL) \cite{cp,lin} parameterisation $w_{\mathrm{CPL}} (z) = w_0 + w_a\left(\frac{z}{1+z}\right)$, which is used in the Dark Energy Task Force report \cite{DETF}. The values of $w_0, w_a$ obtained using this expansion are given below.
\newline

\textit{From Hubble rate measurements}

\begin{eqnarray}
\label{degenhub}
w_0 &=& -\frac{\Omega_k + 3\Omega_{\mathrm{DE}}}{3(1-\Omega_m)}\nonumber\\
w_a &=& \frac{4}{3}\frac{\Omega_k \Omega_{\mathrm{DE}}}{(1-\Omega_m)^2}
\end{eqnarray}

\textit{From luminosity distance measurements}\\
\begin{eqnarray}
\label{degend}
w_0 &=& -\frac{\Omega_k + 3\Omega_{\mathrm{DE}}}{3(1-\Omega_m)}\nonumber\\
w_a &=& -\frac{2}{3}\frac{\Omega_k(\Omega_k - \Omega_{\mathrm{DE}})}{(1-\Omega_m)^2}
\end{eqnarray}

\begin{figure}[htpb!]
     \begin{center}
     \subfigure{
          \includegraphics[width=.45\textwidth]{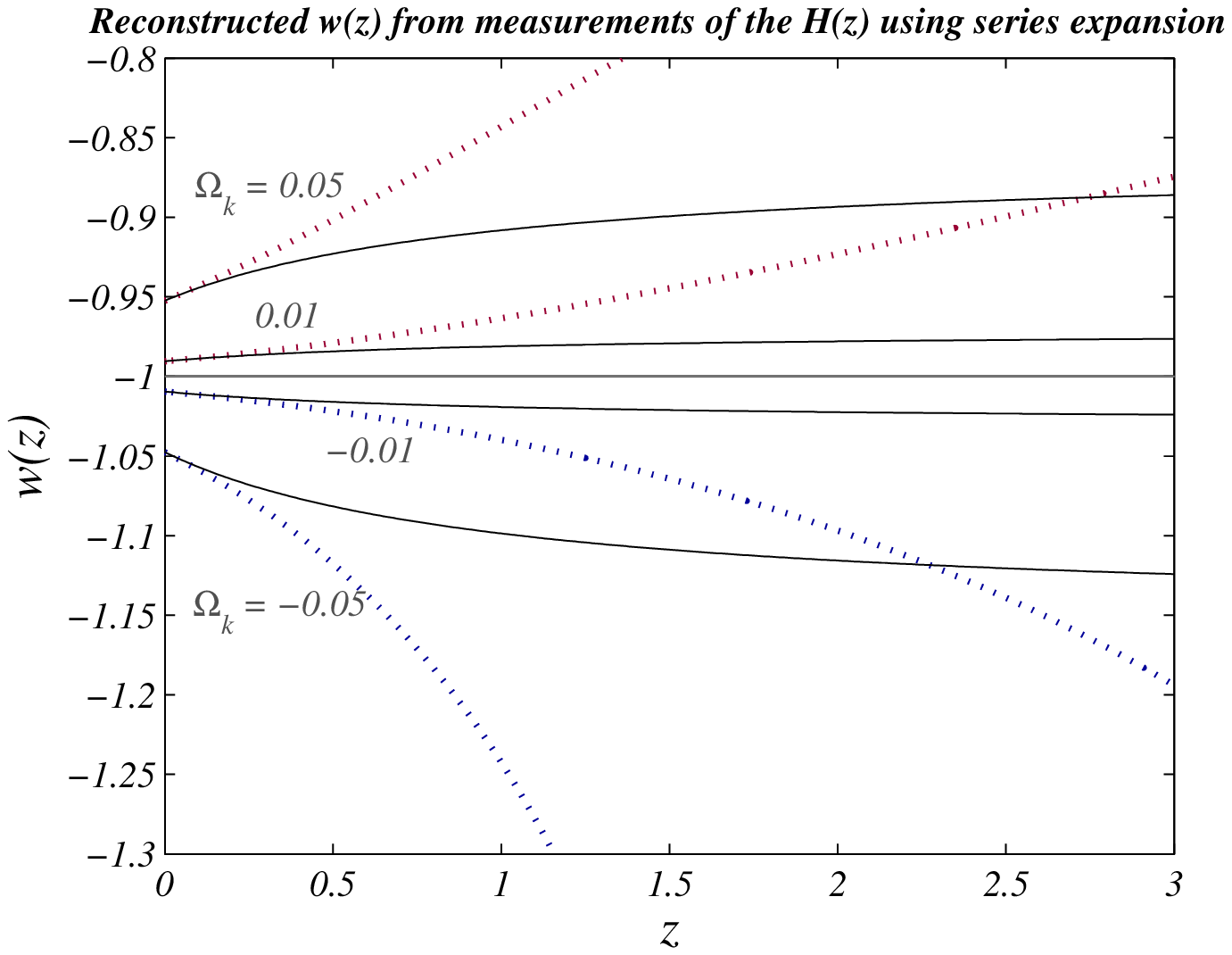}}
     \subfigure{
          \includegraphics[width=.45\textwidth]{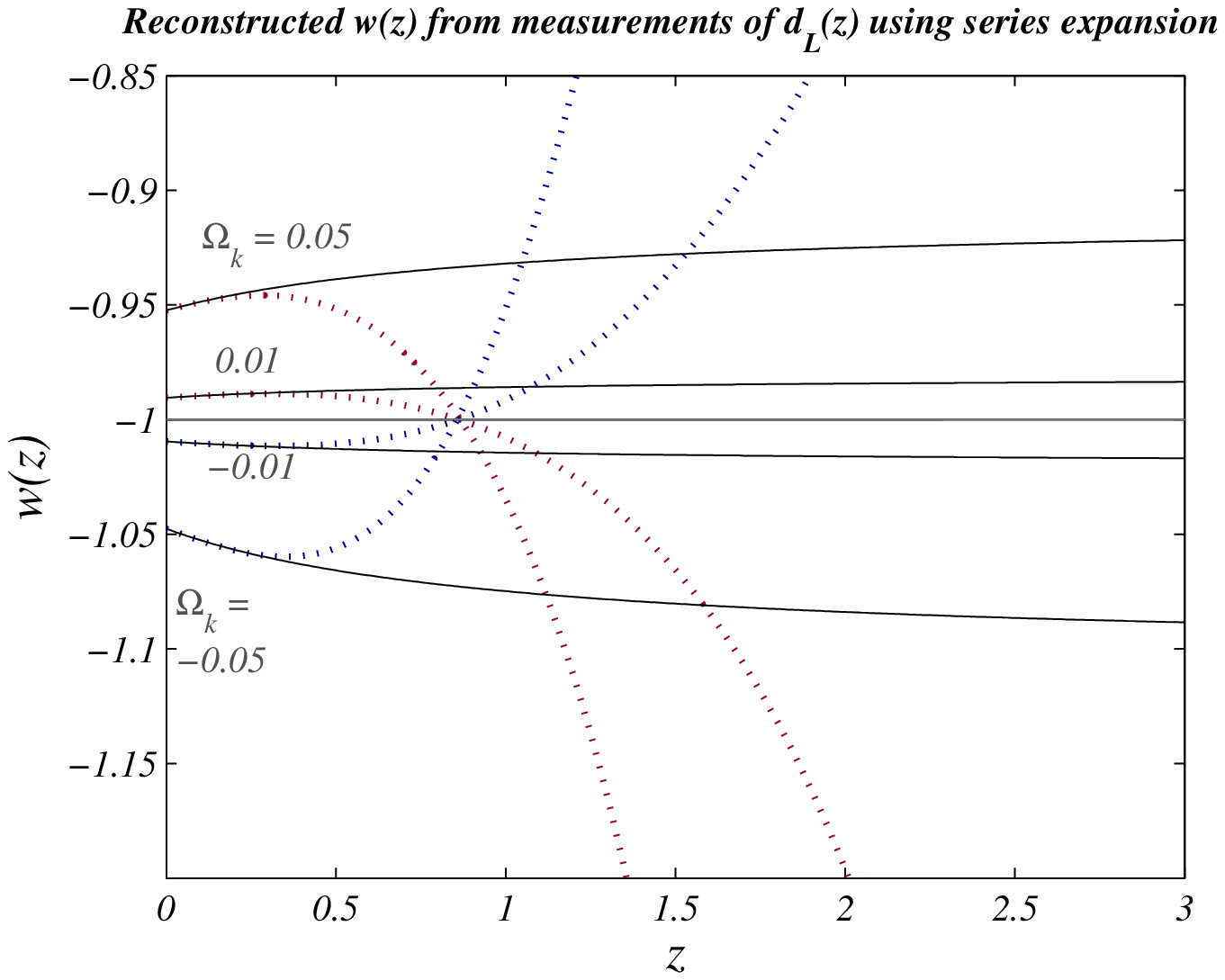}}\\
          \caption{{\bf Degeneracies in standard parameterisations -} $w(z) = w_0 + w_a\frac{z}{1+z}$  using the coefficients in Eqs. (\ref{degend}) and (\ref{degenhub}) (solid lines) compared with the fully non-parametric $w(z)$ inferred from Hubble and distance measurements. Using a limited parameterisation of $w(z)$ like this incorrectly makes it appear that dark energy and curvature are not completely degenerate, leading to artificially strong constraints on curvature and $w_0, w_a$.}
    \label{hi}
	\end{center}
\end{figure}

We plot in Figure \ref{hi} the non-parametric reconstructed $w(z)$ along with the reconstructed $w_{\mathrm CPL}(z)$ from the coefficients given by Eqs.~(\ref{degenhub}, \ref{degend}) for the observables $H(z)$ and $d_L(z)$.
\newline

\section{Conclusions and Outlook \label{conc}}

We have explored the degeneracies between the dark energy equation of state $w(z)$ and cosmic parameters using a non-parametric approach. This means we are able to write down the precise $w(z)$ that will be reconstructed from perfect data if slightly wrong or biased values for the cosmic parameters $\Omega_k, \Omega_m$ are assumed. This is complementary to traditional methods which typically use an aggressive compression of the $w(z)$ function onto a couple of parameters (usually $w_0,w_a$) and then study the degeneracy between these and other cosmic parameters. Our approach is superior in one way however: degeneracies between $w(z)$ and some cosmic parameters such as $\Omega_k$ can appear to be quite weak in the parameterised approach. However, in the case of distance measurements this is completely artificial and due to strong assumptions about the allowed form of $w(z)$ since the degeneracy is perfect if $w(z)$ is allowed to be totally free.

We extend the work of \cite{CCB} to show the reconstructed $w(z)$ from measurements of volume for both wrongly assumed $\Omega_k$ and $\Omega_m$. As with Hubble and distance measurements we show that the errors in $w(z)$ that result from uncertainty in the cosmic parameters are {\em much} larger than the uncertainty in $\Omega_k$ or $\Omega_m$,  especially at large redshifts. We have shown that curvature affects measurements of $H(z)$ and $D(z)$ in complementary ways, with the error at high redshift having opposite signs for an error in $\Omega_k$. In the case of an $\Omega_m$ error, Hubble, distance and volume measurements all lead to the same erroneously reconstructed $w(z)$, a manifestation of the dark matter-dark energy degeneracy highlighted in \cite{m1}.

In this review we have assumed perfect data for Hubble rate, distance and volume at all redshifts. It would be interesting to extend our non-parametric approach to the case of imperfect data which has incomplete redshift coverage and errors on the observables. This is left to future work but will allow contact with the approaches in \cite{arman,pilar}.

{\em Acknowledgments -- }we thank Luca Amendola, Chris Blake, Thomas Buchert, Daniel Eisenstein, George Ellis,
Martin Kunz,  Roy Maartens, Bob Nichol and David Parkinson for useful comments and insights. MC thanks
Andrew Liddle and FTC for support. BB and CC acknowledge support from the NRF and RH acknowledges funding from KAT.


\end{document}